# Vacuum Rabi splitting of a dark plasmonic cavity mode revealed by fast electrons

Ora Bitton[1,5], Satyendra Nath Gupta[2,5], Lothar Houben[1,5], Michal Kvapil [3,4,5], Vlastimil Křápek [3], Tomáš Šikola[3,4] & Gilad Haran [2*]

Recent years have seen a growing interest in strong coupling between plasmons and excitons, as a way to generate new quantum optical testbeds and influence chemical dynamics and reactivity. Strong coupling to bright plasmonic modes has been achieved even with single quantum emitters. Dark plasmonic modes fare better in some applications due to longer lifetimes, but are difficult to probe as they are subradiant. Here, we apply electron energy loss (EEL) spectroscopy to demonstrate that a dark mode of an individual plasmonic bowtie can interact with a small number of quantum emitters, as evidenced by Rabi-split spectra. Coupling strengths of up to 85 meV place the bowtie-emitter devices at the onset of the strong coupling regime. Remarkably, the coupling occurs at the periphery of the bowtie gaps, even while the electron beam probes their center. Our findings pave the way for using EEL spectroscopy to study exciton-plasmon interactions involving non-emissive photonic modes.

[1] Chemical Research Support Department, Weizmann Institute of Science, POB 26, 7610001 Rehovot, Israel. [2] Department of Chemical and Biological Physics, Weizmann Institute of Science, POB 26, 7610001 Rehovot, Israel. [3] Central European Institute of Technology, Brno University of Technology, Purkyňova 123, 612 00 Brno, Czech Republic. [4] Institute of Physical Engineering, Brno University of Technology, Technická 2, 616 69 Brno, Czech Republic. [5]These authors contributed equally: Ora Bitton, Satyendra Nath Gupta, Lothar Houben, Michal Kvapil. *email: gilad.haran@weizmann.ac.il

Cavity quantum electrodynamics deals with the interaction of quantum emitters with the electromagnetic (EM) fields within optical cavities[1]. When this interaction is large enough to reach the so-called strong coupling limit, coherent excitations comprising both cavity modes and emitter modes are generated by excitation with resonant EM radiation. These coherent excitations are manifested by phenomena such as vacuum Rabi splitting, namely the appearance of two branches in optical spectra. Strong coupling is of importance for many quantum technological applications, such as quantum information processing[2,3] and quantum communication[4,5]. In recent years, it has been shown that strong coupling to EM modes can also be used to modify photophysics and chemical reaction dynamics[6,7].

Plasmonic cavities (PCs) are nanometric structures that focus EM fields to sub-diffraction volumes through the effect of surface plasmon excitations[8,9]. The deep sub-diffraction volumes of the EM modes of PCs make them attractive for quantum optical applications. PCs are often made of metallic nanostructures, and rely on the concentration of the EM energy in gaps between particles or at sharp tips or corners. Due to the lossy nature of metals, PCs typically have only moderate quality factors, but their ultra-small mode volumes facilitate strong coupling even under ambient conditions[10]. Indeed, we recently showed that individual PCs can couple to individual or just a few quantum emitters[11]. In particular, we coupled semiconductor quantum dots (QDs) to plasmonic bowties, and observed vacuum Rabi splitting in light scattering spectra. Similar observations were reported more recently by other labs using both scattering and emission spectra measured from individual devices[12,13].

PCs can sustain a set of multipolar resonances, some of which are bright while others are dark and subradiant[8]. It is typically the lowest energy, bright dipolar resonance that is used for coupling to quantum emitters. However, dark modes might also be of significant interest for quantum optical studies, particularly since they are expected to have longer lifetimes than bright modes and therefore should be able to store EM energy for longer times[14,15]. It is difficult, though, to excite and probe dark modes of individual plasmonic structures, due to their inability to couple radiatively to the far-field.

Electron energy loss (EEL) spectroscopy is an electron microscopy method that can access the EM near-field and yield the complete spectrum of plasmonic modes in nanostructures with a very high spatial resolution[16]. EEL spectroscopy has been applied extensively to study both bright and dark resonances of various plasmonic structures[17–20]. It probes the out-of-plane component of the electric field around a plasmonic device[16], and the EEL signal is closely related to the optical extinction spectrum of a plasmonic structure[16,21,22].

Here, we show that EEL spectroscopy in a scanning transmission electron microscope (STEM) can be used to probe the coupling of silver bowtie PCs with QDs residing in their vicinity. We demonstrate vacuum Rabi splitting in EEL spectra due to coupling between the QDs and the dark mode of the PCs. In contrast to expectation, the strongest coupling occurs at the periphery of the PC gaps. EM simulations allow us to attribute this finding to the non-uniform distribution of the EM field of the dark mode.

## Results

**Fabrication of plasmonic bowties coupled to QDs.** We fabricated silver bowties using electron beam lithography on $SiO_2$ membranes, 18 nm thick, and inserted QDs into their gaps and around them. For more details, see Methods section.

Figure 1a shows an elemental map of such a device, obtained by energy dispersive x-ray spectroscopy (EDS). The bowtie is protected by a thin alumina layer against oxidation and the CdSe/ZnS QDs are seen around it.

**Bright and dark plasmonic modes of a bare bowtie revealed by EEL spectra.** The EEL spectrum measured from a PC depends on the position of the electron beam, as different positions excite a different combination of plasmonic modes[18]. Figure 1b shows EEL spectra measured at two different positions on a bare PC devoid of QDs. When the electron beam interacts with the outer edge of the bowtie (orange dot in the inset of Fig. 1b), the lowest-energy excited plasmonic mode is found at 1.8 eV (Fig. 1b, orange spectrum). From the calculated charge density map obtained at this frequency (Fig. 1c), it is clear that this is the lowest-energy bright longitudinal dipolar mode of the PC. In contrast, when the electron beam passes through the center of the PC gap (red dot in the inset of Fig. 1b), a mode at 2.1 eV is excited, whose charge distribution identifies it as the lowest-energy dark mode. This mode is formed by an antiparallel combination of the dipolar plasmons of the individual prisms of the bowtie, carries a zero net dipole moment (Fig. 1c)[18] and is subradiant[23]. Additional modes excited with the electron beam at other positions are shown in Supplementary Figs. 1 and 2, and will not be further discussed below.

Figure 1d, e shows EEL maps at certain energy slices, each corresponding to one of the peaks shown in Fig. 1b. Thus, for example, the map at energies between 1.76 and 1.85 eV corresponds to the bright dipolar mode discussed above. Interestingly, the EEL signal, which (as noted above) is sensitive to the out-of-plane component of the EM field, is stronger on the outer edge than in the gap. On the other hand, for the energy slice around 2.1 eV (corresponding to the dark dipolar mode), the out-of-plane field is much more confined in the gap region than at the outer edges.

**EELS spectra of bright and dark plasmonic modes coupled to QDs.** When we introduce into PC gaps QDs whose lowest optical transition (the so-called band-edge exciton[24]) is in resonance with the bright dipolar mode, EEL spectra show a clear evidence of a Rabi splitting. Two such spectra are presented in Fig. 2b, d, and ADF-STEM images of the devices are shown in Fig. 2a, c. The Rabi splitting values in these devices are 200 and 150 meV, respectively. The splitting is an indication of the formation of plasmon–exciton polaritons, new states that mix (to a different degree, depending on the coupling strength) the exciton and plasmon states. To obtain the coupling strength characterizing each spectrum, we fitted the spectra to the coupled oscillator model[25]. Since, as already noted, it has been shown that the EELS signal is closely related to the optical extinction spectrum[16,21], we used the expression for extinction derived from the coupled oscillator model (see Methods)[25]. We find coupling strengths of $105 \pm 2$ meV and $68 \pm 3$ meV. Similar coupling strengths were reported in our previous work and in others' for a small number of QDs[11–13]. An EM simulation of the EELS signal using the boundary-element method (BEM)[26,27] with four QDs positioned in the PC gap yields a similar splitting (120 meV) in the lowest energy mode (Fig. 2e, f).

Remarkably, when QDs with a band-edge exciton at 2 eV are introduced into PC gaps, EEL spectra measured with the beam within the gaps are also found to split (Fig. 3), demonstrating that the QDs can also couple to the dark dipolar mode. The three panels of Fig. 3 demonstrate coupling to the dark mode, with Rabi splitting values of 160, 150 and 100 meV. Fits to the coupled oscillator model give values of 83 ± 2, 75 ± 2 and 57 ± 1 meV, respectively, for the coupling strength.

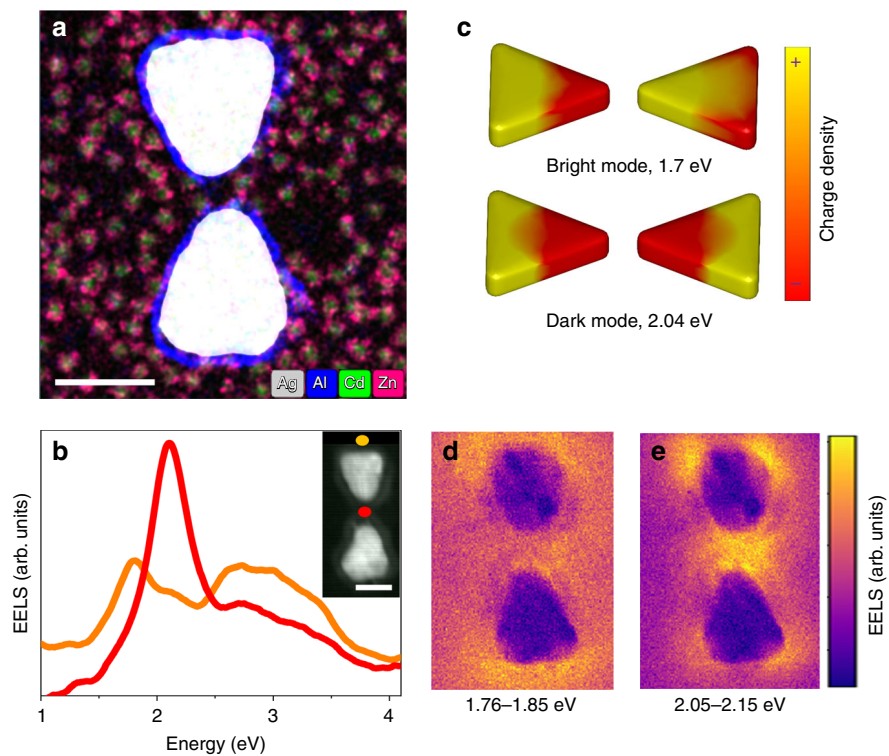

**Fig. 1 Plasmonic bowties coupled with quantum dots. a** Electron micrograph of a plasmonic bowtie with QDs in its vicinity. The image is a color-coded superposition of elemental maps taken by EDS in the STEM. The color key is shown at the bottom of the image. The scale bar represents 50 nm. **b** EEL spectra of a bare bowtie measured at the two locations indicated by matching colored dots in the inset, which is an annular dark-field (ADF) STEM image of the device. The scale bar represents 50 nm. **c** Simulated charge-density maps for the bright and dark dipolar modes at the indicated energies. **d**, **e** Experimental EEL maps for the bright (**d**) and dark (**e**) modes, constructed at the indicated energy ranges.

**Coupling takes place at the periphery of the bowtie gap.** In order to understand the mechanism of the exciton–plasmon interaction involving the dark mode, we performed BEM simulations of the EEL spectra of coupled devices. Unexpectedly, with four and even eight QDs positioned in the PC gap as above, no splitting was observed in the simulated spectrum (Fig. 4a). However, when ten additional QDs were added at the periphery of the PC gap rather than its center (Fig. 4b), a clear splitting was observed in the spectrum, implying that the QDs at the periphery contribute more to the exciton–dark mode interaction than the ones at the center. Indeed, a simulation with QDs positioned only at the gap periphery (Fig. 4c) showed a similar splitting. To shed more light on this surprising observation, we plot a map of the simulated total electric field distribution around a bare PC at 2 eV, shown in Fig. 4d as a projection on the $x$–$y$ plane. It is clearly seen that the electric field at the center of the gap, and even near the tips of the two prisms forming the bowtie, is very small. On the other hand, a very high field is observed at the periphery of the gap, comparable to the highest field in the bright mode map (Supplementary Fig. 2a). Since the coupling strength is linearly proportional to the electric field, it is expected that QDs that reside at the center of the gap will not couple at all with the dark dipolar mode, while QDs at the periphery will strongly couple with it. Notably, the out-of-plane component of the electric field has a small contribution in the gap (Fig. 4e, showing a projection on the $x$–$y$ plane), ~50 smaller than the total field, but this small contribution is enough to couple to the electron beam and generate an EEL signal.

To verify that indeed coupling at the periphery of the gap is responsible for the observed Rabi splitting, we looked for devices in which the QDs were not located at the center of the gap. Figure 3a shows an example of such a situation; even while the electron beam probes the center of the device, there are no QDs there, and the QDs situated at the periphery couple to the PC, leading to the appearance of Rabi splitting in the EEL spectrum. This observation directly confirms the simulation results. An additional such example is shown in Supplementary Fig. 3.

**Discussion**
We reported here vacuum Rabi splitting in EEL spectra that can be traced to coupling between quantum emitters and a dark plasmonic mode of an individual PC. This interaction cannot be accessed optically in the far field, as the dark mode is not radiant. However, the ability to excite the dark mode with an electron beam opens the way to observe the respective PC–QD interaction in EEL spectra. EEL spectroscopy has been used recently to probe strong coupling to bright plasmonic modes of individual plasmonic structures[28,29]. While the merit of dark modes for strong coupling phenomena has been discussed theoretically[30,31], we are not aware of any previous experimental study of the coupling to a dark mode at the level of a single PC. (One recent study exploited the coupling of a dark lattice resonance mode of a plasmonic array to a layer of molecules for lasing[32].) The coupling effect depends on the local distribution of the EM field in the PC gap region. Counterintuitively, it is not the gap center that provides the largest coupling, even though this is exactly where the combined excitations are probed.

It is instructive to ask whether the coupling strength values we observe here put our PC–QD systems in the strong coupling regime. To that end, we compare our measured $g$ values to two criteria often discussed in the literature[33]. The first criterion, $g > (\gamma_p - \gamma_e)/4$ (where $g$, $\gamma_p$ and $\gamma_e$ are the coupling strength, plasmon linewidth and exciton linewidth, respectively), guarantees two real solutions in the coupled-oscillator model when the QD is resonantly tuned to the plasmon and may be seen as defining a lower bound for the strong-coupling regime. Put in a different way from the quantum mechanical point of view, when this criterion is fulfilled, the system has passed an exceptional point[34,35] and is guaranteed to possess two distinct eigenstates. Based on the values of $\gamma_p$ and $\gamma_e$ we measure in our systems, the above criterion gives a value of ~55 meV, and the values of $g$ extracted from our spectra (up to 85 meV) are larger.

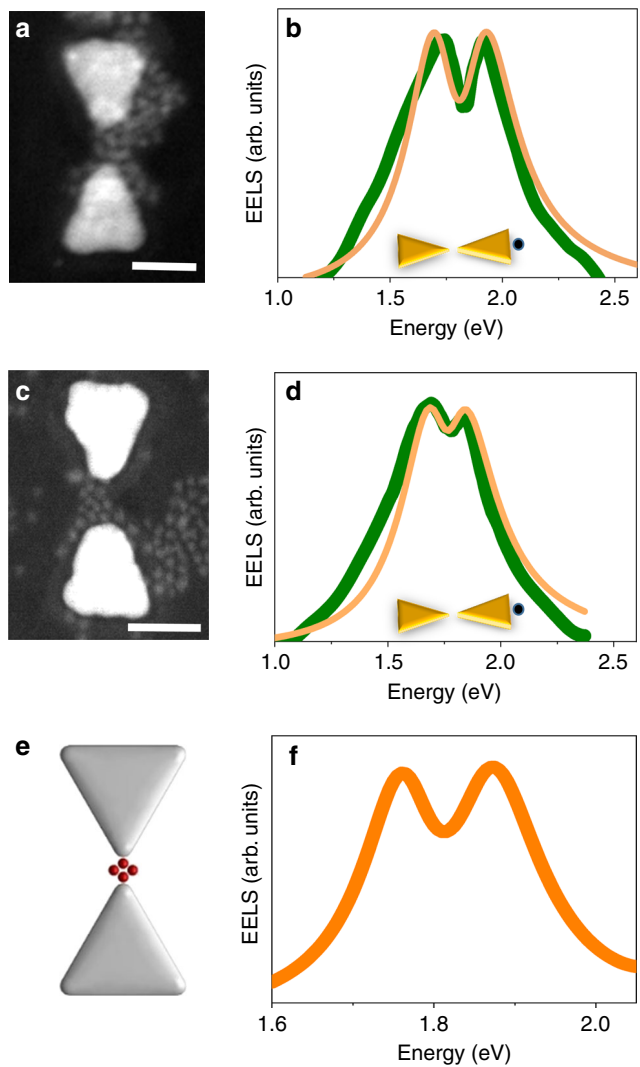

**Fig. 2 Coupling with the bright dipolar mode. a, c** ADF-STEM images of two devices loaded with QDs with an exciton energy of 1.8 eV. The scale bars represent 50 nm. **b, d** EEL spectra of the bright dipolar mode of the devices in **a** and **c**, demonstrating Rabi splitting. Green curves are experimental data, and orange curves are fits to the coupled oscillator model, as described in the Methods section. Insets demonstrate the points of excitation by the electron beam. **f** BEM simulation of the spectrum of the bright mode in a PC coupled to four QDs, as shown in **e**.

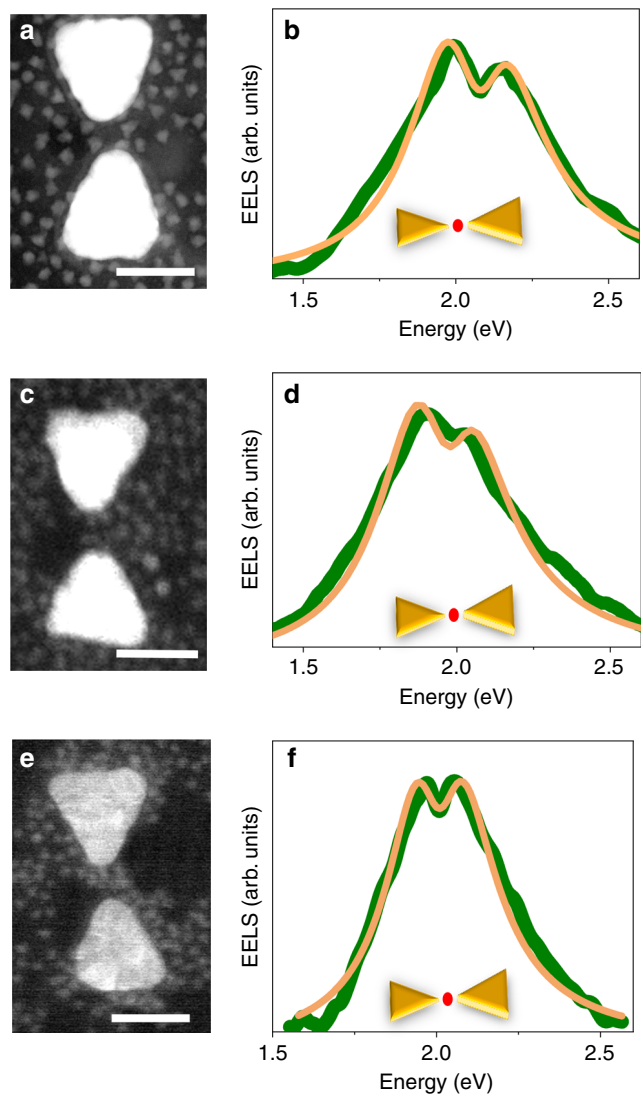

**Fig. 3 Coupling with the dark dipolar mode. a, c, e** ADF-STEM images of three devices loaded with QDs with an exciton energy of 2.0 eV. The scale bars represent 50 nm. **b, d, f** EEL spectra of the dark mode of the devices in a, c and e demonstrating Rabi splitting, as with the bright mode. Green curves are experimental data, and orange curves are fits to the coupled oscillator model, as described in the Methods section. Insets demonstrate the points of excitation by the electron beam.

While this criterion guarantees two eigenstates, it does not ensure that measured spectra will clearly show these states as separate peaks. Therefore, a second and stricter criterion is introduced. This criterion[33], given by $g > (\gamma_p + \gamma_e)/4$, is more heuristic, and requires a larger $g$ to be fulfilled. Indeed, in our case it gives a value of ~120 meV, larger than the values we report here, although not by a lot. However, in reality, the splitting of the two peaks in the spectrum grows continuously, as demonstrated by simulated spectra in Supplementary Fig. 4. The spectra are clearly split already when $g = 55$ meV (i.e. the value of the first criterion), and the splitting increases as $g$ grows. Indeed, splitting is readily observed in the experimental spectra (Figs. 2 and 3). Therefore, we believe we can safely state that all the PC–QD systems measured here are found at the onset of the strong-coupling regime.

As noted already in the introductory remarks of this paper, the appeal of using dark (excitonic or plasmonic) modes rather than bright modes for quantum optical applications resides with their longer lifetimes. For example, a dark exciton of a self-assembled QD has been used as a qubit that emits clusters of entangled photons[36]. The lifetime of a dark plasmonic mode should be increased due to the lack of radiative damping, and this longer lifetime should affect the dynamics of coupling between a PC and quantum emitters[31]. Somewhat surprisingly, the longer lifetime does not lead to a significantly smaller linewidth of the dark mode peak in our EEL spectra (compared to the bright mode, Fig. 1). This might point to factors that dephase the excitation of the dark mode and therefore contribute to its linewidth significantly without affecting its lifetime. This possibility likely requires time-dependent studies with femtosecond resolution to be tested. Future experimental work might be able to probe the relation between the long lifetime of the dark mode and Rabi oscillations in the coupled PC–QD system. If the involvement of emissive QDs makes the dark plasmon–exciton polaritons partially emissive, this phenomenology might be detected in the far field using time-resolved cathodoluminescence[37].

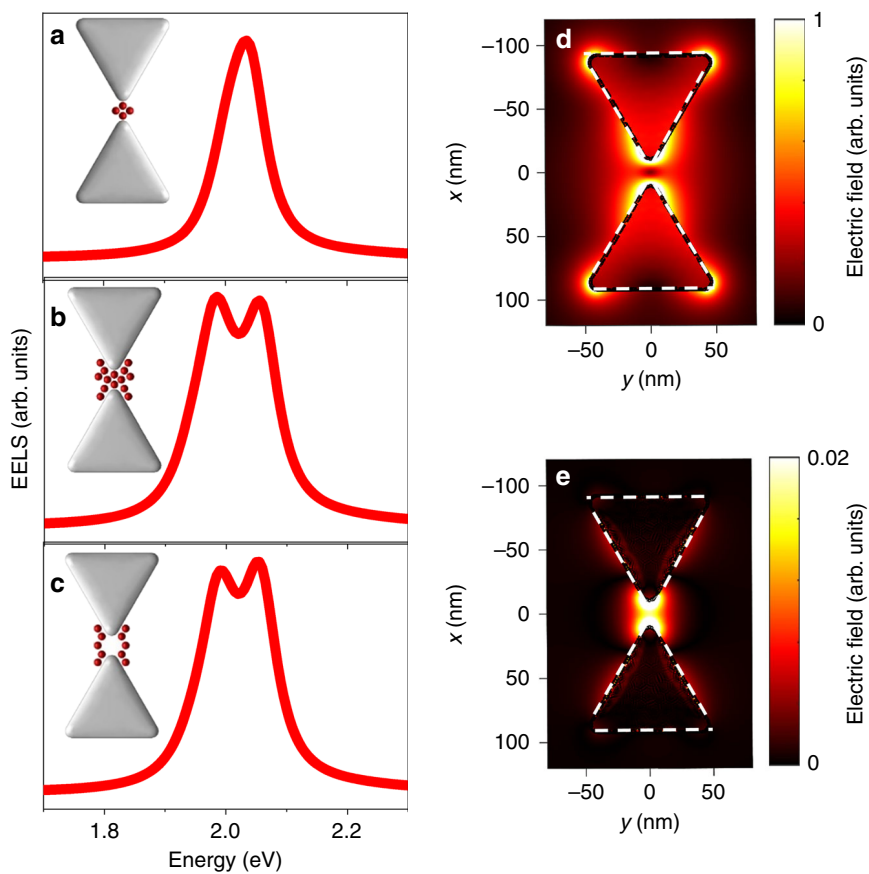

**Fig. 4 EM simulations of dark mode coupling with the QDs. a** Dark mode spectrum when four QDs are positioned in the center of the gap. **b** Dark mode spectrum when ten additional dots are added at the periphery of the gap. **c** Dark mode spectrum with QDs positioned only at the gap periphery. **d** Map of total electric field distribution around the PC at 2 eV, projected on the *x*–*y* plane. **e** Map of the out-of-plane electric field distribution around the PC at 2 eV, projected on the *x*–*y* plane. The maps in **d** and **e** were calculated for bowties without QDs. Note their different color scales.

## Methods

**Materials**. Electron microscopy grids with 18-nm-thick SiO$_2$ suspended films were purchased from Ted Pella Inc. A poly(methyl methacrylate) 950K A2 (PMMA) electron beam resist, used for the fabrication of bowties, was obtained from MicroChem, as was the solvent stripper Remover PG. Chromium (used as an adhesive layer) and silver were obtained from Kurt J. Lesker. Water-soluble mercapto-undecanoic acid capped CdSe/ZnS core/shell nanocrystals (QDs) were acquired from MK Impex Corp. Thrimethylaluminum (TMA), used as a precursor for atomic layer deposition, was purchased from Strem Chemicals Inc.

**Electron beam lithography and fabrication of silver bowties**. PMMA was spin-coated on the grids at a speed of 5000 rpm for 50 s to achieve a thickness of 60 nm, followed by baking at 180 °C for 90 s. The PMMA-coated grids were loaded into a Raith E_line Plus electron-beam lithography system and the PMMA was exposed to define the shape of bowties. The accelerating voltage used for the exposure was 30 kV and the beam current was 30 pA. The design consisted of matrices of bowties, with each matrix hosting 12 bowties. Each bowtie was separated by 10 μm from its neighbors. To remove the exposed PMMA, substrates were developed using 1:3 methyl isobutyl ketone:isopropyl alcohol for 30 s, followed by immersion in a stopper (isopropyl alcohol) for 30 s and drying with a nitrogen flow. An electron-beam evaporator (Odem) was used for metal deposition on the patterned substrates. Initially, chromium was evaporated to deposit a 2-nm adhesion layer, followed by silver. The lengths of the silver prisms generated in this way ranged from 65 to 85 nm, and their thickness was 25 nm. After metallization, a lift-off step was performed in a solvent stripper for 40 min to obtain silver bowties on the SiO$_2$ films. To protect the bowties from oxidation, we deposited on the grid a conformal thin layer of alumina (2.5 nm) using an atomic layer deposition system (Fiji Plasma), with TMA as a precursor. Finally, PMMA was again spin coated on the grid in order to protect the bowtie structure from degradation during the incorporation of QDs, as the alumina layer was found not to be sufficient in protecting the bowties within a water-based solution.

**Incorporation of QDs into bowtie structures**. We first made the PMMA surface hydrophilic by applying a delicate N$_2$ plasma using an inductively coupled plasma apparatus (Plasma Therm Ltd.). We then immersed the sample vertically in a water-based solution of QDs and allowed the solution to evaporate within a low vacuum environment. During the evaporation, QDs were randomly positioned on the PMMA surface. In the last stage, we immersed the grid in acetone to dissolve the resist. Under the conditions we used (unheated remover and a thin resist), the QDs were not completely washed away with the resist, and the majority of the QDs tended to randomly deposit on the substrate, sticking in the vicinity of the bowties.

**EEL measurements**. Electron microscopy data were recorded using a double aberration-corrected Themis Z microscope (Thermo Fisher Scientific Electron Microscopy Solutions, Hillsboro, USA) equipped with a high-brightness Schottky field emission gun and a Wien-type monochromator at an accelerating voltage of 80 kV. EDS hyperspectral data were obtained with a Super-X G2 four-segment SDD detector with a probe semi-convergence angle of 30 mrad and beam current of ~200 pA. Low-loss EEL maps were recorded on a Gatan Quantum GIF 966ERS energy loss spectrometer (Gatan Inc., Pleasanton, USA) with an Ultrascan1000

CCD camera. The low-loss EEL maps were recorded using a monochromated STEM probe with an energy width of 70–80 meV, a semi-convergence angle of 34 mrad and a beam current of 160 pA. The outer semi-collection angle of the spectrometer was set to 18 mrad.

Data analysis was performed in the software DigitalMicrograph (Gatan Inc., Pleasanton, USA). In all figures, the zero-loss peak was removed by subtracting its part to the left of the y axis, reflected to the right side of the axis, from the whole spectrum. In Figs. 2 and 3, the spectra of specific plasmonic modes were isolated by first subtracting a background signal measured far from the plasmonic structures and then substracting the contribution of neighboring modes after fitting them to Gaussian curves. Raw spectra were smoothed using a Savitzky–Golay filter.

**EM simulations**. Numerical simulations of EEL maps and spectra were performed using the boundary-element method as implemented in the Matlab toolbox MNPBEM[26]. A numerical model of a bowtie was constructed from two equilateral triangular prisms with an edge length of 73 or 87 nm, a height of 20 nm, radii of curvature at the vertices of 7 nm, and a gap size of 20 nm. The edge lengths were selected to match the experimental resonance frequency for the bright and dark modes. The complex refractive index of silver from Johnson and Christy[38] was used, and a refractive index of 1.35 was assumed for the ambient medium. The bowtie was excited by an electron beam (energy 80 keV) at normal incidence. Simulations of EEL spectra for a bowtie loaded with QDs were performed with dots modeled as spheres of 8 nm located within the gap and its periphery. The complex dielectric function of these dots was approximated by a Lorentz model (similarly to our previous work[11]) with a high-frequency dielectric constant $\epsilon_\infty = 6.0$, $\omega_{0,bright} = 1.75$ eV, $\omega_{0,dark} = 1.99$ eV, $\gamma_0 = 0.08$ eV, and an oscillator strength $f = 0.8$.

**Spectral fitting with the coupled oscillator model**. EEL spectra were fitted with the expression for the extinction of two coupled oscillators at frequency $\omega$[25]:

$$S(\omega) \propto \omega \, \mathrm{Im}\left( \frac{\left(\omega_e^2 - \omega^2 - i\sqrt{(\gamma_e^2 + \delta^2)}\omega\right)}{\left(\omega^2 - \omega_p^2 + i\sqrt{(\gamma_p^2 + \delta^2)}\omega\right)\left(\omega^2 - \omega_e^2 + i\sqrt{(\gamma_e^2 + \delta^2)}\omega\right) - 4\omega^2 g^2} \right), \quad (1)$$

where $\omega_e$, $\omega_p$, $\gamma_e$ and $\gamma_p$ are the resonance frequencies and linewidths of the emitter and cavity, respectively, and $g$ is the coupling constant. We fixed the values of $\gamma_e$ (110–130 meV) and $\gamma_p$ (350–395 meV), based on the linewidths of the photo-luminescence spectra of individual QDs and the scattering spectra of empty bowties (it was found that these linewidths slightly varied between devices). We broadened the linewidths of the emitter and the cavity by the instrumental broadening $\delta$ (70 meV).

### Data availability

All relevant data are available from the authors upon request.

### Acknowledgements
G.H. is the incumbent of the Hilda Pomeraniec Memorial Professorial Chair. This work was partially supported by EU project 810626 – SINNCE.

### Author contributions
O.B., S.N.G and G.H. designed the research. O.B., S.N.G. and L.H. performed the experiments. M.K., V.K. and T.S. performed EM simulations. All authors participated in writing the manuscript.

### Competing interests
The authors declare no competing interests.

### Additional information

**Supplementary information** is available for this paper at https://doi.org/10.1038/s41467-020-14364-3.

**Correspondence** and requests for materials should be addressed to G.H.

**Peer review information** *Nature Communications* thanks Matthew Pelton and the other, anonymous, reviewer(s) for their contribution to the peer review of this work.

**Reprints and permission information** is available at http://www.nature.com/reprints

**Publisher's note** Springer Nature remains neutral with regard to jurisdictional claims in published maps and institutional affiliations.

Supplementary Information

# Vacuum Rabi splitting of a dark plasmonic cavity mode revealed by fast electrons

Bitton et al.

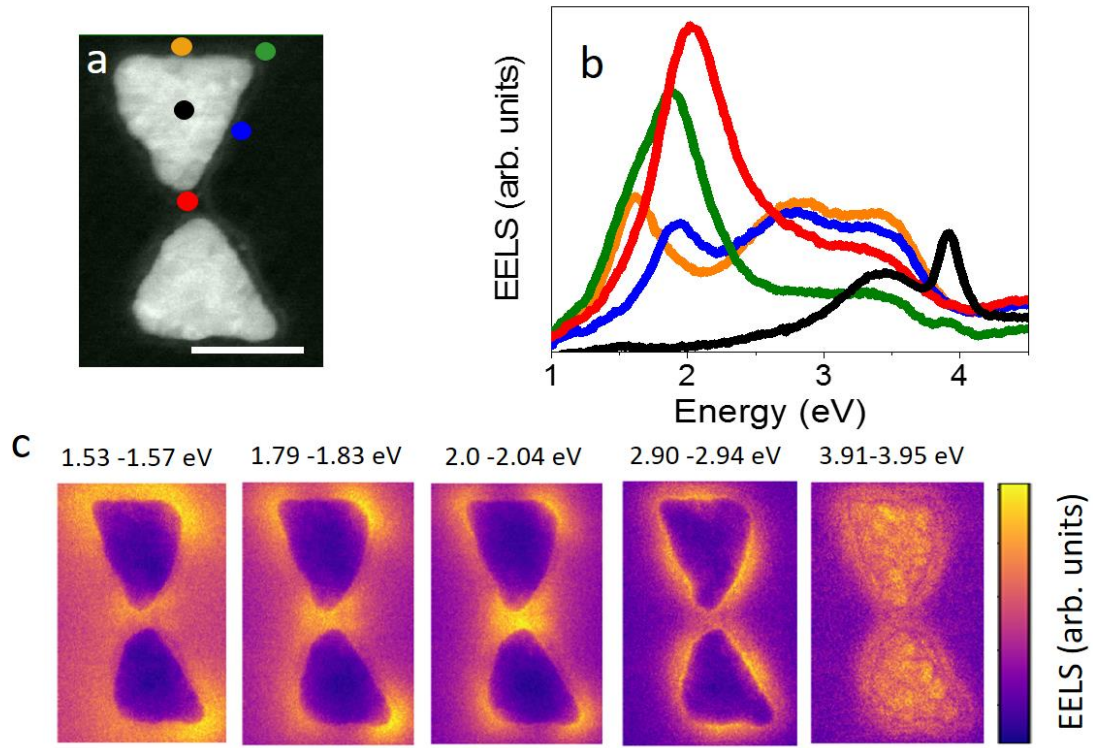

**Supplementary Fig 1. EEL spectroscopy of a bare plasmonic bowtie.** (a) STEM image of a bowtie without QDs. Scale bar is 50nm. (b) experimental EEL spectra obtained for different beam locations indicated by matching colored dots in a. (c) Experimental EEL maps for different specified energies ranges, corresponding to the different modes obtained in b.

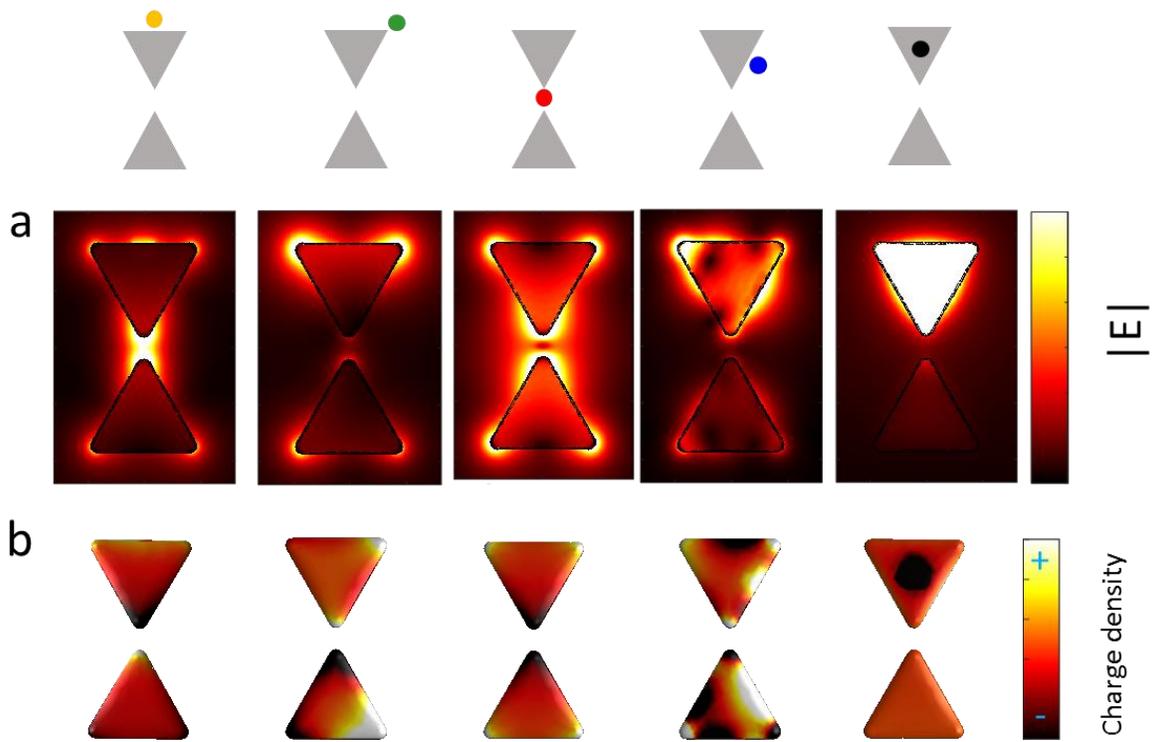

**Supplementary Fig 2. Simulated EEL maps for the different modes.** (a) Total electric field distribution projected on the x-y plane. (b) Charge density. All Maps were created for a bare PC at a certain energy and beam location, as specified in the upper part of the figure.

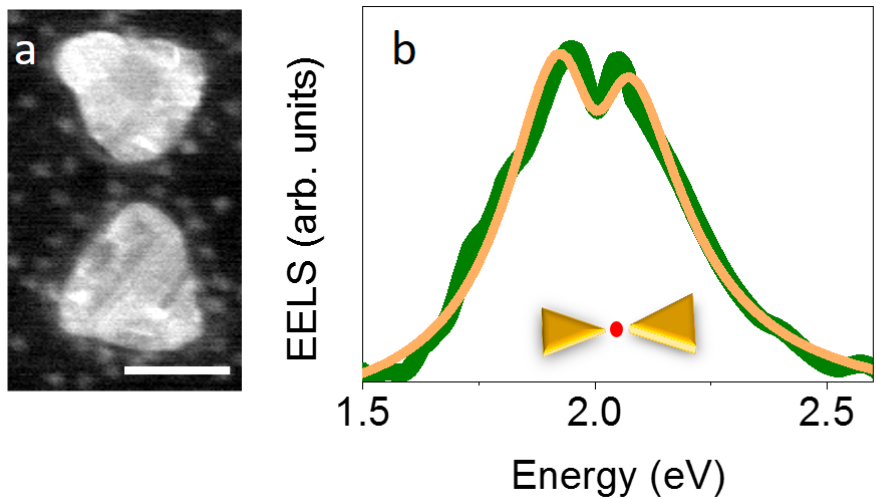

**Supplementary Fig. 3. A second example of QDs coupling to a PC at the periphery of the gap.** (a) STEM image of a device loaded with QDs with an exciton energy of 2.0 eV such that all emitters are located away from the center of the gap. Scale bar represents 50 nm. (b) EEL spectrum of the dark dipolar mode of the device in a showing Rabi splitting. The green curve is experimental data, while the orange curve is a fit to the coupled oscillator model, as described in the Methods section. The fit yields a coupling strength of 64±1 meV. Inset demonstrates the point of excitation by the electron beam.

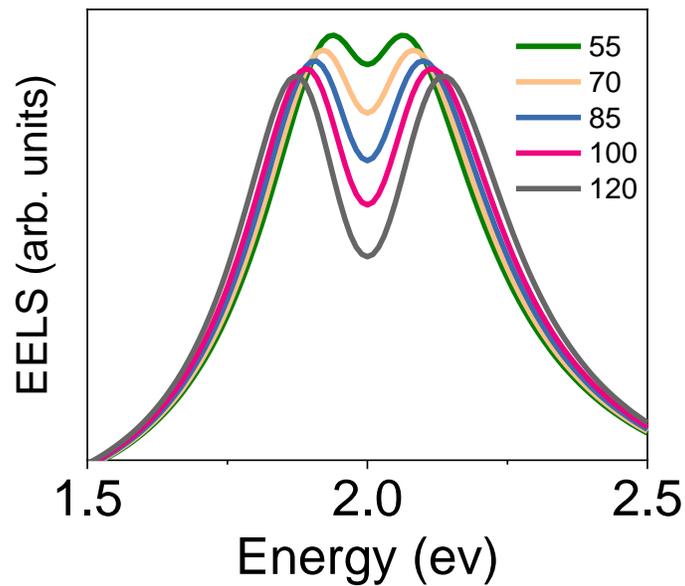

**Supplementary Fig. 4. The continuous nature of Rabi splitting.** Coupled-oscillator model simulations of EELS spectra with increasing coupling strengths, given in the legend in meV. A clear splitting is already seen with a coupling strength of 55 meV, though the two branches of the spectrum split significantly when the second criterion discussed in the text is fulfilled (120 meV). The largest dark-mode coupling strength in our EEL spectra is 85 meV (blue spectrum).